# REPEATING THE EÖTVÖS-PEKÁR-FEKETE EQUIVALENCE PRINCIPLE MEASUREMENTS*

Gábor Péter
BME [1], Department of Control Engineering and Information Technology
László Deák
MTA[2], Wigner Research Centre for Physics
Gyula Gróf
BME, Department of Energy Engineering
Bálint Kiss
BME, Department of Control Engineering and Information Technology
György Szondy
Independent Researcher
Gyula Tóth
BME, Department of Geodesy and Surveying
Péter Ván
MTA, Wigner Research Centre for Physics
Lajos Völgyesi
BME, Department of Geodesy and Surveying

Loránd Eötvös, with his colleagues Dezső Pekár and Jenő Fekete from 1906 [1] carried out a series of measurements (EPF measurements) validating the proportionality of the gravitational and inertial masses.

In 1986, Fischbach and his colleagues revealed systematic deviations in the measurements. These deviations were slightly above the threshold of error, and were neither reproduced nor explained by later measurements, which were much more precise although they used a different method.

As a result of our analysis of the EPF measurements, we found a possibility of systematic error [2] that may justify repeating the experiments under the improved conditions offered by today's state-of-the-art technical facilities. The measurements are planned to be carried out in the Jánossy Underground Research Laboratory at a depth of 30 m in the KFKI site, under appropriate undisturbed and controlled conditions. The repetition of the EPF measurements will be particularly relevant in the Eötvös Year 2019, which is planned to commemorate the 100th anniversary of the death of Loránd Eötvös. In this paper, we briefly review the background, preparations, and current status of the measurements.

## Background

Between 1906 and 1908, Loránd Eötvös and his colleagues performed measurements with an Eötvös torsion balance to check the equivalence of the gravitational and the inertial mass. The measurements were based on the assumption that the force of gravity, the weight of the test body, is the combination of the gravitational force towards the centre of the Earth and the centrifugal force perpendicular to the Earth's axis of rotation. And if the gravitational force material dependent, then the direction of the resulting gravitational force will also depend on the material, which can be detected by an east-west oriented Eötvös torsion balance.

The measurements showed only negligible, seemingly random, deviations, a few times of the $10^{-9}$ error margin.

Later, in 1986, Fischbach et al. found that these small deviations were not entirely random but could be described as a linear function of the binding energy of the nuclei, and proposed the existence of a short-range so-called fifth force [3, 4, 5]. The hypothesis has been the subject of considerable debate. Eötvös' results were analysed in detail, and new, far more precise measurements were made, but they did not reveal the predicted Yukawa-like, finite-range fifth force. This was all the more expected because Eötvös and his colleagues checked the effect on one pair of materials in the Sun's gravitational field and found similar differences [1], so the observed deviations are presumably not related to distance at that scale.

We had no doubts about the equivalence of equivalence of gravitational and inertial mass at long distances, because several other measurements have been made since Eötvös' time. Before Fischbach's, Robert Dicke and his colleagues had verified the equivalence principle with

---

an accuracy of $10^{-11}$, and Braginsky et al. with an accuracy of $10^{-12}$ [7, 8].

Robert Dicke and colleagues introduced several innovations. First, using a north-south oriented torsion balance, they measured the matter dependence of the relationship between the Sun's gravitational attraction and the centrifugal force due to the Earth's orbit. The advantage of this was that the pendulum did not need to be rotated during the measurement because of the 24-hour periodicity of the signal. In addition, a triangular pendulum has been created, which is less sensitive to gradient effects. Moreover, the pendulum, consisting of a gold-aluminium material pair, was placed in a vacuum and damped and balanced by two electrodes via a feedback loop. The constant drift of the quartz fibre used was also compensated electrostatically. From the data recorded over several days of measurement cycles, a computer was used to statistically filter out disturbance signals caused by temperature fluctuations and mechanical vibrations (construction work taking place about 30 m from the measurement site).

Braginsky et al. measured similarly to Dicke and his group. One method of greatly increasing the sensitivity was to increase the length of the torsion wire made of tungsten to an extreme length of about 3m.

After 1986, the emergence of the fifth force has given a new boost to these experiments. Since 1990, similar, increasingly precise measurements have been carried out by the Eöt-Wash group (University of Washington). In their latest and most precise torsion balance measurements, they have achieved an accuracy of $10^{-13}$ [5]. A speciality of these researches is that they measure at extreme gradients to test Yukawa interactions up to 1 m in range. The resulting gradient effects are carefully compensated. The group has also performed measurements where the Sun and the Milky Way are considered as attractors [9].

The Lunar Laser Ranging (LLR) experiment compares the Sun's gravitational attraction to the Earth and the Moon by accurately measuring the Earth-Moon distance. The research showed that the gravitational acceleration of the two celestial bodies is the same within a measurement error of $10^{-13}$. This result has several implications. If the gravitational self-interaction causes a change in the ratio of gravitational to inertial mass - this would violate the strong equivalence principle. This type of "binding energy" is many orders of magnitude smaller than the binding energy in the nuclei, hence the experiment proved the strong equivalence principle to an accuracy of $(2.3\pm3.2)\times10^{-4}$.

This measurement assumes that the composition of the Earth and the Moon is significantly different, because the Earth has an iron-nickel core similar to that of iron meteorites, while the Moon does not [9]. However, the geochemical-meteorite model assuming an iron-nickel core is incompatible with our current knowledge of geophysics, because it cannot explain, for example, the disappearance of shear stresses in the outer core. A more physically realistic model is the so-called astrophysical model, according to which there is no significant difference in the material composition of the Earth's mantle and core, the difference being due to phase transition caused by high pressure [10]. However, because of the Moon's small mass, the Moon's interior does not develop enough pressure to create an Earth-like phase transition, and therefore has neither a core nor an Earth-like magnetic field. The rock samples from the Moon show no significant difference in the material composition of the two bodies, and their ages are identical, suggesting a simultaneous origin from the same material source and the same material structure [11]. Thus, the LLR experiment is probably not suitable to test the weak equivalence principle.

The MICROSCOPE experiment, which has been collecting data since December 2016, plans to improve the accuracy of previous research by a further two orders of magnitude, using the difference in relative acceleration of test masses orbiting the Earth in slightly different orbits. The planned "Galileo Galilei", with its torsional pendulum designed for zero gravity in space, would improve the accuracy of the weak equivalence principle by two orders of magnitude and verify the validity of the weak equivalence principle with an accuracy of $10^{-17}$ [12].

Nevertheless, the EPF measurements were not reproduced in the original form and with the original instruments, and after more recent measurements using more sophisticated methods, the reason for the systematic discrepancy observed in the Eötvös work was ignored [6].

According to our hypothesis, the systematic error in the Eötvös measurements is due to a higher order effect of the gravity field gradient caused by a force depending on the shape of the samples used. Repeating the EPF experiment with attention to this effect may not only confirm the above-mentioned assumption of a systematic error, but may also contribute new perspectives to the current most accurate ground measurements.

Plans

Our preliminary studies so far have delivered two major results from the perspective of the measurement. One is that the accuracy can be improved by at least one and a half orders of magnitude with today's technology. The other is that while the original EPF measurements could have been affected by environmental effects (e.g. temperature and gravity gradients) of unknown magnitude and in unknown ways (the exact magnitude of which we can only guess at in the absence of original data), the new measurements we plan could be performed under controlled conditions of both temperature and environmental gradients. According to preliminary calculations the environmental effects at the proposed measurement site are minimal and the gravity gradients are well controlled and do not appear to be too large.

For this reason, it would be important to perform EPF measurements using the original method but with the higher precision provided by today's technology, because this would allow to clearly determine the cause of the correlation that Fischbach and his colleagues expected to discover in Eötvös' results, and to analyse the role of the gradient effect in the current ground and satellite-based equivalence principle experiments.

On 23 June 2017, the decision was made at the Wigner Research Centre for Physics to repeat and reproduce the EPF measurements with the Eötvös torsion balance, equipped with modern instruments, and to find the real cause of any discrepancies.

The project, initiated by the Society for the Unity of Science and Technology (ETTE), will be carried out between June 2017 and June 2019 in cooperation with the Wigner Research Centre for Physics, the Department of Geodesy and Surveying, the Department of Control Engineering and Information Technology of Budapest University of Technology and Economics (BME), and other departments, organisations and experts.

To achieve this objective, the following main tasks have been defined:
1. Full automation of the Eötvös torsion balance measurements, with the possibility of remote control.
2. Increase the accuracy of the torsion balance using modern techniques and equipment.
3. To replicate EPF measurements in a maximum controlled and documented environment.
4. Thorough evaluation of the measurements and publication of the results.

## Preparations so far

Two different types of Eötvös torsion balances are available for the planned measurements: an Eötvös-Rybár Auterbal (Automatic Eötvös-Rybár Balance) instrument and an Eötvös-Pekár torsion balance. The Auterbal torsion balance (shown in Figure 1) was developed in the 1920s by István Rybár, successor of Eötvös, at the Department of Experimental Physics. The most significant improvements compared to the earlier models, besides the reduction of the detection time to 40 minutes per azimuth, were the rotation of the instrument by means of a clock mechanics and the photographic recording of the readings [13]. The automatic reading allowed the unattended operation of the instrument, but the frequent failures of the fragile mechanics required constant attention.

In the case of the development of the Eötvös-Pekár balance shown in Figure 2, Dezső Pekár focused on reducing the size and swing time while maintaining the simplicity of the instrumentation and insisting on a more accurate and reliable visual reading. The torsion balances developed by Pekár are known as Eötvös-Pekár balances, but their official type designation was Small original Eötvös G-2 [13]. Three versions of the instrument were manufactured, differing basically only in the length of the torsion fiber. In the instruments manufactured in l926 the length of the fiber was still 50 cm, in the 1928 instruments it was 40 cm, and in the 1930 instruments it was only 30 cm.

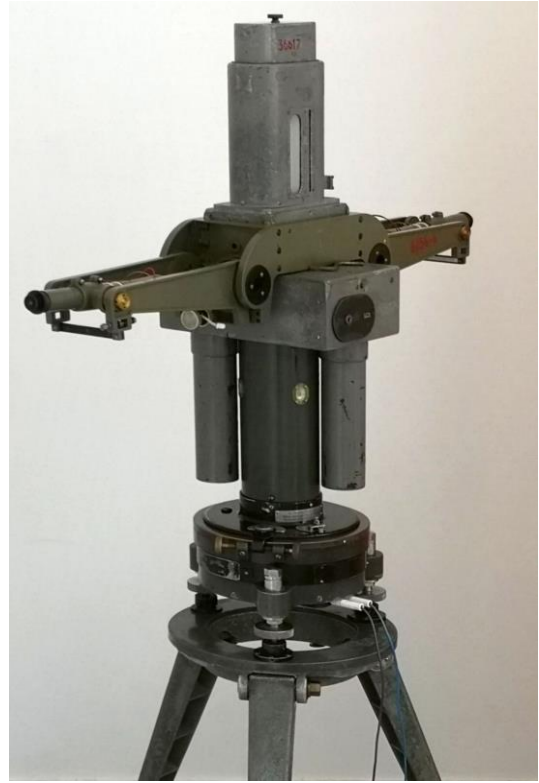

*Figure 1.* The repaired Eötvös-Rybár (Auterbal) torsion balance.

Since the available instruments were used a very long time ago, the first important step was to renovate, repair, check their usability, to calibrate [14] and to adjust them. This took many months. In order to minimise the torsional drift of the fibres, they were continuously loaded for more than half a year in the unlocked position of the torsion balances. By loading the fibres over a long period of time, both fibres of the Pekár balance have reached a nearly drift-free state, while only one torsion fibre of the Auterbal balance has reached a nearly drift-free state. Of course, this makes the Auterbal balance suitable for measurements, as the drift can be easily compensated during the evaluation of the measurements, but for EPF measurements with extreme precision requirements it is more convenient to use the Pekár torsion balance. In addition, since the easy replacement of the masses in the instrument is a decisive factor for the repetition of EPF measurements, we have clearly decided to use the 30 cm fibre length Pekar pendulum for EPF measurements and will concentrate further developments on this instrument only.

The most important requirement for successful measurements is to eliminate disturbances to measurement as much as possible. The most dangerous sources of error for EPF measurements are: the mass of the person reading the instrument, vibrations of the balance, temperature variations and high gravity gradients around the instrument. The latter problem can be easily addressed by choosing a suitable location and by accurately surveying the Eötvös tensor elements in the instrument environment.

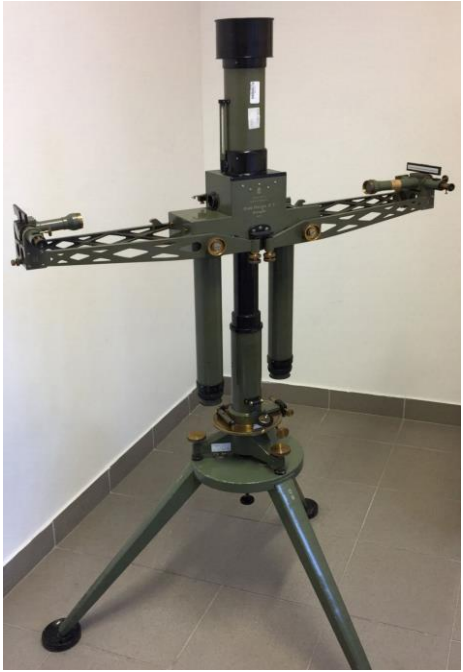

*Figure 2.* The Eötvös-Pekár torsion balance.

By selecting a suitable site, the stability of the environmental temperature and especially the ground vibrations caused by traffic can be eliminated. Therefore, we chose the Jánossy Underground Research Laboratory at 30m depth as the site for the experiments. According to our preliminary measurements, the thermal stability of the site is adequate, the temperature variation is only a few hundredths of a degree per day and the expected microseismic ground vibration is sufficiently low, which will be checked by seismometers.

The most significant and dangerous source of error is mainly the mass effect, the presence of the person handling and reading the instrument [15]. However, the presence of the person also disturbs the thermal equilibrium and causes clearly perceptible ground vibrations and even, according to experience, the small tilting of the instrument. The source of the error can be avoided by eliminating the direct human presence: fully automating and remotely controlling the measurement process. This requires the use of digital instrument read-outs instead of visual read-outs, using appropriate CCD sensors and computer image processing, and the use of a remote-controlled motor to rotate the instrument to different measurement azimuths.

By mounting a CCD sensor in place of the ocular of the Pekar torsion balance for visual readout, the position of the torsion balance arm can be determined by the value of the digital image produced (Figure 3). According to our experiments, the electronically captured image using a CCD camera of appropriate resolution takes up a 25-30 pixel per division of the scale. A proper image processing technique can be used to determine the position corresponding to a fraction of a pixel, so that the expected reading accuracy from an image is more than 1/100th of a scale division ($10^{-10}$). This is already more than one order of magnitude better than the $10^{-9}$ accuracy of the original visual detection. Add to this the fact that by taking several images of the position of the balance every second, up to 1000 images can be captured and evaluated in a few minutes, and this can result in a further increase in reading accuracy.

An important question in the measurement is: is the digital image processing fast and efficient enough? The relative scale shift between two consecutive images can be determined quickly and accurately by image correlation using the original scale. The automatic determination of the absolute position of the scale, i.e. the complete reading, is much more difficult and requires the use of artificial intelligence, since the numbers and their positions in the scale must be recognised.

Therefore, we designed a new barcode scale and evaluation procedure that makes the computer reading much simpler and more efficient, while at the same time increasing the accuracy of the reading. Bar-coded length measurement technology has been successfully used for decades by various geodesic instrumentation companies for high accuracy, customized digital leveling. By analogy, we have designed a barcode scale from the visible part of the CCD that can be used to unambiguously reconstruct the distance to the scale zero point, i.e. the scale reading of interest, with high accuracy in any reading situation. As the bar-coded scale is not suitable for conventional visual detection, we designed a combination of bar-coded and conventional numbered scales, as shown in the upper graph in Figure 3. This preserves the possibility of traditional detection, while allowing for an ideal and fast barcode evaluation for computer reading. The combined scales are being produced by laser engraving, and if their accuracy meets our needs, they will be used for the measurement. Unfortunately, the quality of the first barcode scales did not meet our expectations by far, so we had to learn more about the production techniques and adapt the new designs to them.

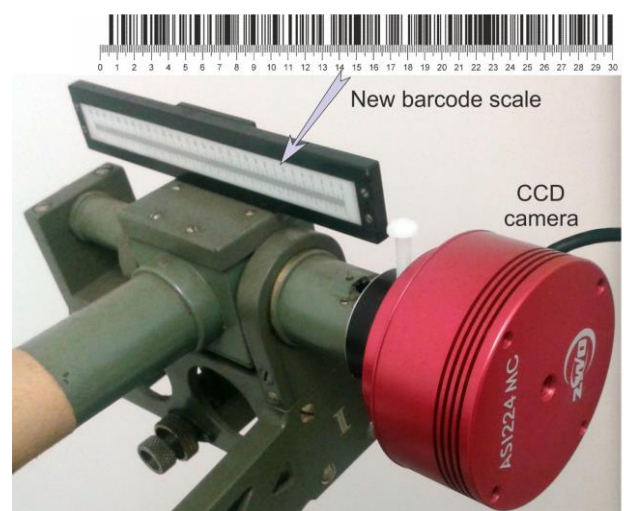

*Figure 3.* CCD sensor mounted at the place of the ocular.

The most critical element of the mechatronic solutions is the implementation of a remote-controlled instrument rotation unit. An important requirement for this unit is to avoid permanent magnetic disturbances caused by the rotating motor and its associated structure during measurements at rest between rotations. The control of the motor had to be such that the rotation of the instrument in azimuths 0°, 90°, 180°, 270° was accurate and reliable, without sudden accelerations and decelerations, and without mechanical shock to the very sensitive structure of the torsion balance. During rotation, correct in- and out-travel and stopping can also have a positive effect on the damping of the instrument's oscillation, shortening the time required for damping.

To rotate the torsion balance into different measurement azimuths, the apparatus shown in Figure 4 was constructed, rotated by a special motor via a ribbed belt connected to the free-turning upper part of the instrument. Most of the components are 3D printed. The current reduction ratio is 1:10, with a 10 and a 100 tooth wheel on the motor and pendulum side respectively. The angular position is measured with the appropriate resolution by an optically actuated encoder disc and optocouplers on the motor-side axis of the transmission. The disc has 100 lines, which, combined with the double phototransistor design, gives a total of 400 increments every turn. Since the counting of the increments is done on the motor side axis, for the full 360° range of the torsion balance, the resolution of axis is 0.09°.

If during tests the gear ratio does not appear to be satisfactory, an additional worm gear can be added to the system to further reduce the torsional fibre stresses. This will of course improve the accuracy of the angular position measurement as well.

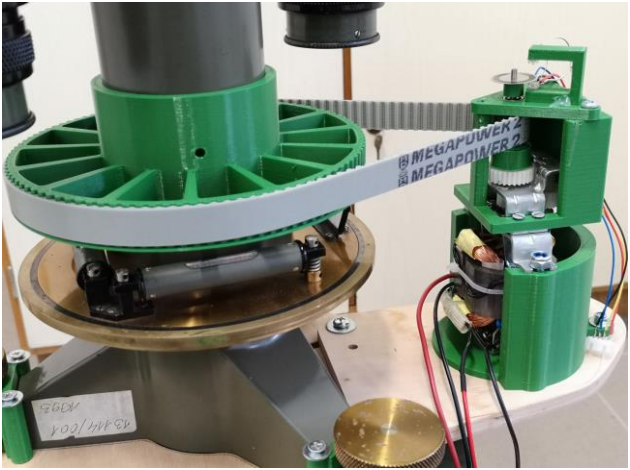

*Figure 4.* Solution for remote-controlled instrument rotation.

A schematic of the measurement automation solution architecture is shown in Figure 5. The embedded microcontroller unit (RTdev) is responsible for real-time data acquisition from sensors placed near the torsion balance (temperature, humidity, etc., on-demand), power supply monitoring and control of the instrument's motion unit, including the angular position transmitters. Non time-critical operations are realized on a personal computer (PChost). This is also where the rotation sequences, which can be defined for each individual measurement, are executed for automated measurement, where the data acquisition on the CCD sensors is controlled, and where the data and images are logged, saved and processed. PChost also receives camera images. No real-time processing is required, but the exact time of image capture must be recorded.

An important requirement for successful measurements is the supply and manufacture of precision special parts, in addition to the unique IT solutions required for modernization of the instrument. The next most difficult task is to produce test masses of the right shape and weight and to hang them on the appropriate fibres.

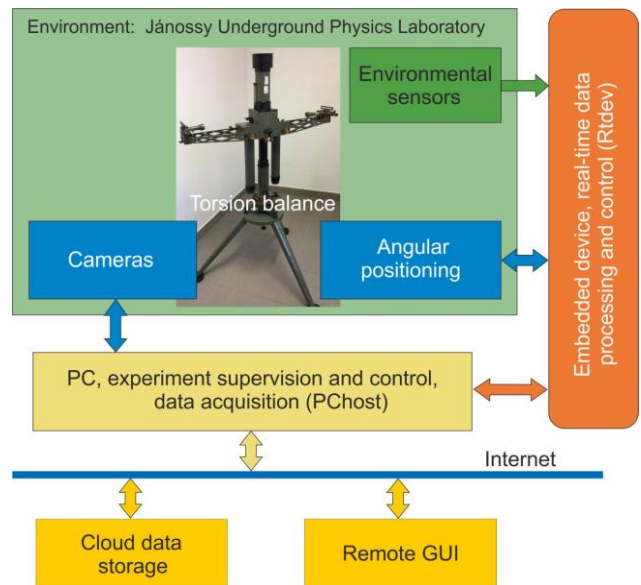

*Figure 7.* The data acquisition control system

## Summary

Loránd Eötvös and his colleagues performed a series of experiments on the equivalence of gravitational and inertial masses, and obtained small deviations, just above the error threshold. After a careful study of the relevant publications, we found a potential source of systematic error in the measurements that justifies repeating the experiments under improved conditions offered by today's modern technology.

In June 2017, the Wigner Research Centre for Physics re decided to repeat the EPF measurements with the Eötvös torsion balance and to investigate the real cause of any violations. The Wigner Research Centre for Physics, the BME Department of Geodesy and Surveying, the Society for the Unity of Science and Technology (ETTE), the BME Department of Control Engineering and Information Technology, and other organisations, departments and experts are involved in the preparations for the measurements.

The measurements are planned to be executed at a depth of 30 m in the Jánossy Underground Research Laboratory at KFKI, under appropriate quiet and controlled conditions. Preparations are in progress, the renovation of the Jánossy laboratory is about to be completed, and we are currently working on the modification of the Eötvös-Pekár torsion balance for measurements at the BME Department of Geodesy and Surveying. The preliminary calibration measurements and the installation of the control motor have been done, the installation of the CCD sensors and the development of the control software, the manufacturing of the barcode scale and the test mass are planned or in progress. We plan to start the measurements in December 2018.

The main instrument of the measurements, the Eötvös-Pekár torsion balance, is owned by the Institute of Geodesy and Geophysics of the Hungarian Academy of Sciences. We would like to express our thanks to the management of the institute for making the instrument available for our measurements. We also thank Fornax 2002 Kft. and Pál Sári for their technical assistance.

## Authors

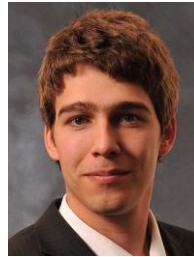

*Gábor Péter* graduated as an electrical engineer in 2015 from the Department of Control Engineering and Information Technology at BME, where he is currently a PhD student. His research topic is mapping unknown domains with cooperating agents.

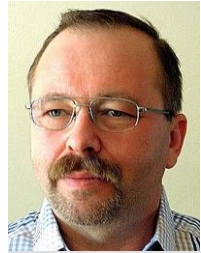

*László Deák* is a physicist and senior research fellow at the Nuclear Materials Department of the Institute of Particle and Nuclear Physics of the Wigner Research Centre for Physics. His research interests include X-ray neutron and Mössbauer spectroscopy, their applications in specular and diffuse scattering.

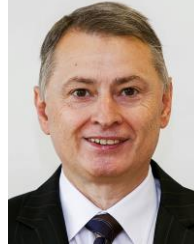

*Gyula Gróf,* mathematician-engineer, Head of the Department of Energy Engineering Faculty of Mechanical Engineering, BME. His main research interests are experimental and theoretical investigations of energy and transport processes, development of thermophysical measurements. He has built special scanners and photo-robots, which have been used in geological research in the area of Bátaapáti.

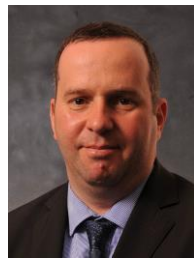

*Bálint Kiss*, electrical engineer (96'), head of the Department of Control Engineering and Information Technology, associate professor. His research interests are robotics and control of mechatronical systems.

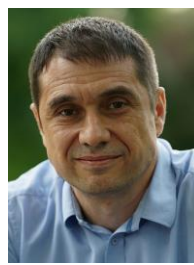

*György Szondy* graduated as an electrical engineer in 1994. Since then he has been working on IT systems development. He is currently running an IT startup. As an independent researcher he has been actively involved in fundamental, theoretical and practical issues of gravity for 18 years. Since 2006 he has been a board member of the Association for the Unity of Science and Technology (ETTE).

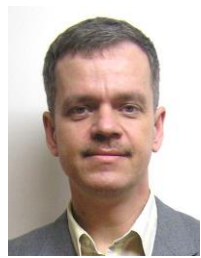

*Gyula Tóth*, associate professor, candidate of engineering, graduated in surveying in 1985. Since then he has been teaching and researching at the Department of Geodesy and Surveying of the Technical University of Budapest. His research interests include physical and geomathematical geodesy, in particular the geodetic determination of the geomatical shape of the Earth, the geoid. He was awarded the Academic Prize for his research in this field in 2011.

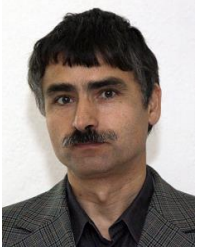
*Péter Ván* is a physicist and senior research fellow at the Particle and Nuclear Physics, Department of Theoretical Physics of the MTA Wigner Research Centre for Physics and the Department of Energy Engineering, at the Faculty of Mechanical Engineering, BME. His research interests is non-equilibrium thermodynamics. In the last years he has been coordinating the Wigner FK research at the Laboratory of Gravitation and Geophysics in the Mátra.

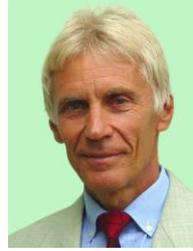
*Lajos Völgyesi,* geophysicist, Doctor of the Hungarian Academy of Sciences. Professor Emeritus at the Department of Geodesy and Surveying of the BME. His main research interests are physical, astronomical and mathematical geodesy, measurement and investigation of the Earth's rotation and gravity, further development and applications of the Eötvös torsion-balance. He has been awarded the Academy Prize and is chairman and member of several national and international scientific committees.